\documentclass[11pt]{article}

\pdfoutput=1
\usepackage{arxiv}
\usepackage{hyperref}
\usepackage[normalem]{ulem}
\usepackage{amsthm}
\usepackage{amsmath}
\usepackage{amssymb}
\usepackage{mathrsfs}
\usepackage{mathtools}
\usepackage{graphicx}
\usepackage{enumerate}
\usepackage{comment}
\usepackage{algorithm}
\usepackage{algpseudocode}
\usepackage{color}
\usepackage{stmaryrd}
\usepackage{booktabs} 
\usepackage{fullpage}
\usepackage{appendix}
\usepackage[table]{xcolor}

\theoremstyle{definition}
 
\theoremstyle{remark}

\theoremstyle{remark}

\numberwithin{theorem}{section}

\title{Detecting Patterns of Interaction in Temporal Hypergraphs via Edge Clustering}

\author{
Ryan DeWolfe\thanks{Department of Mathematics, Toronto Metropolitan University, Toronto, Canada; e-mail: \texttt{ryan.dewolfe@torontomu.ca}}
\And
Fran\c{c}ois Th\'{e}berge\thanks{Tutte Institute for Mathematics and Computing, Ottawa, Canada; email: \texttt{theberge@ieee.org}}
}

\begin{document}

\maketitle

\begin{abstract}
Finding densely connected subsets of vertices in an unsupervised setting, called clustering or community detection, is one of the fundamental problems in network science.
The edge clustering approach instead detects communities by clustering the edges of the graph and then assigning a vertex to a community if it has at least one edge in that community, thereby allowing for overlapping clusters of vertices.
We apply the idea behind edge clustering to temporal hypergraphs, an extension of a graph where a single edge can contain any number of vertices and each edge has a timestamp.
Extending to hypergraphs allows for many different patterns of interaction between edges, and by defining a suitable structural similarity function, our edge clustering algorithm can find clusters of these patterns.
We test the algorithm with three structural similarity functions on a large collaboration hypergraph, and find intuitive cluster structures that could prove useful for downstream tasks.
\keywords{Edge Clustering, Hypergraphs, Line Graphs, Dynamic Graph Clustering}
\end{abstract}

\section{Introduction}
One of the main topics in unsupervised data science is clustering: finding groups of similar data that match intuition. 
In the context of complex networks~\cite{miningcomplexnetworks}, clustering (also called community detection) algorithms often try to partition the vertices so that there are many edges within communities and few edges between communities.
Unfortunately, there is not a single definition of what makes a community, which allows for a myriad of community detection algorithms~\cite{fortunatoreview}.

One approach to community detection is called \textit{edge clustering} \cite{edgeclustering1, edgeclustering2, hierarchicaledgeclustering, scaleabledynamicedgeclustering}.
As the name implies, these algorithms cluster the edges of a graph.
A vertex is then assigned to a community if at least one of its incident edges is in that community.
An immediate benefit of shifting the perspective to the edges is that we can allow for community overlap and hierarchical structure~\cite{hierarchicaledgeclustering}.
Edge clustering algorithms perform clustering on the \textit{line graph} instead of the original graph.
For a graph $G$, the vertices of its line graph $LG$ correspond to the edges in $G$, i.e. $V(LG) = E(G)$, and there is an edge between two vertices in $LG$ if the corresponding edges are adjacent in $G$.
In a simple undirected graph, adjacent edges always share exactly one node, so many algorithms weight the line graph using additional information to assess the similarity of adjacent edges.
For example, in \cite{hierarchicaledgeclustering}, the authors measure edge similarity with the Jaccard index of the closed neighbourhood of the distinct nodes.
Edge clustering then proceeds by applying a standard node clustering algorithm to the line graph to obtain clusters of edges.

However, a graph is not always the best data structure for modelling interactions that take place between more than two entities.
For example, in collaboration networks~\cite{collaborationnetworks}, vertices may correspond to an author and there is an edge between two authors if they co-author a paper.
Since papers are often written by more than two authors, it is more appropriate to consider a collaboration hypernetwork.
A hypergraph (or hypernetwork) is an extension of a graph where a single edge can include any subset of the vertices (usually restricted to non-empty subsets).
There has been significant recent interest in hypernetwork science \cite{beyondpairwise, hypernetworkwalks} and several proposed methods for community detection \cite{hypergraphmodularity, kumar, mimax, hypergrapcd}, although the added complexity often makes extending traditional network science techniques a non-trivial task.

We can further enrich the representation of collaboration networks by considering timestamps on the hyperedges, such as the date a given paper was written, when they are available.
This leads to a \textit{temporal hypergraph} (also called a dynamic hypernetwork).
There are many existing methods for dynamic community detection \cite{dynamicdetectionreview}, although these methods generally produce a clustering of vertices that evolves through time instead of a single edge clustering that is temporally aware.

In this paper, we apply edge clustering to undirected temporal hypergraphs to find patterns of interaction that are both structurally and temporally similar.
Our method is heavily inspired by the approach described in \cite{scaleabledynamicedgeclustering} for a dynamic directed graph; they suggested extending to a dynamic hypergraph was suggested in Section 6.3 of that paper.
In our work, we explore a few similarity functions and illustrate the usefulness of this approach by analyzing a large collaboration hypernetwork.

The rest of the paper is organized as follows.
In Section \ref{sec:method}, we present the temporal hyperedge clustering algorithm using a Jaccard based structural similarity measure.
We describe a large collaboration hypernetwork in Section \ref{sec:data}, and analyze the results of the algorithm in Section \ref{sec:results}.
Finally, we describe some ideas for future research in Section \ref{sec:conclusion}.

\section{Method}\label{sec:method}
Like a graph, a hypergraph $H$ consists of a vertex set $V$, an edge set $E$, and an incidence function $M$.
The edge set $E$ is an index set, and the incidence function $M:E \to \mathcal{P}(V)$ maps an edge $e$ to the subset of vertices in that edge.
In this notation, a hypergraph is a graph if the output $M$ always has size $2$ (we exclude self-loops).
When considering a simple graph, it is common to refer to the edge as its membership set (for example, if there is an edge between vertices $x$ and $y$ we would write $xy \in E$), but explicitly using an incidence function improves clarity for multi-edges, and, in our case, timestamps.
The temporal information is given as a function on the edges $t:E \to \mathbb{R}$.
Carefully defining the range and scale of the time function is problem specific, but for most cases it is practical to set the value of the earliest edge to $0$ and let $t(e)$ be the number of seconds/days/months/years between the timestamps of edge $e$ and the earliest edge.

As mentioned in the introduction, we are interested in finding clusters of hyperedges that contain similar vertices and happen at similar points in time.
To quantify these similarities, we must make two choices: a hyperedge similarity function $s: E \times E \to [0,1]$ and a time kernel $T: E \times E \to [0,1]$.
The hyperedge similarity function should evaluate the similarity between two sets of nodes, with identical sets having maximal similarity.
A simple option is the Jaccard index: for two hyperedges $i,j \in E$, it is defined as
$$s_{Jac}(i, j) = \frac{|M(i) \cap M(j)|}{|M(i) \cup M(j)|}.$$
We suggest and experiment with two other structural similarity functions in Section \ref{sec:other_s}.
The time kernel defines the similarity between two times.
It is important to set a maximum difference that produces a non-zero time similarity in order to keep the line graph sparse and the algorithm efficient.
We call this maximum difference $\sigma$, and let it be an input to the algorithm.
For $T$, we suggest a simple linear difference that is normalized so that identical times have a similarity of $1$, and two times that are further apart than $\sigma$ have $0$ similarity.
Refer to Section \ref{sec:time_kernel_experiments} for further analysis of the parameter $\sigma$.
For edges $i$ and $j$, the time similarity is given by
$$T_\sigma(i,j) = \max \left\{1 - \frac{|t(i) - t(j)|}{\sigma}, 0 \right\}.$$

We can now construct the weighted line graph.
Define a weighted graph $LG$ where each vertex corresponds to a hyperedge, i.e. $V(LG) = E(H)$, and the weight of an edge $ij \in E(LG)$ is computed by combining the membership and time similarities.
Define the weight function $w:E(LG) \to [0,1]$ as
$$w(ij) = \sqrt{s_{Jac}(i, j) \cdot T_\sigma(i, j)}.$$

Finally, we must apply a method to cluster the vertices of the weighted line graph.
We follow~\cite{hierarchicaledgeclustering} and use single linkage clustering, but instead of optimizing a cut of the dendogram we apply the \textit{excess of mass} cluster extraction method used in HDBSCAN~\cite{hdbscan} (by default we use a  minimum cluster size of $10$).
This method selects a set of non-overlapping clusters from the dendogram in order to maximize a stability score.
While the clusters may not overlap, not every vertex is necessarily in a cluster.
The vertices not in any cluster are called \textit{outliers}.

We use sparse matrices from Scipy~\cite{scipy} to construct the weighted line graph $LG$ and a maximum spanning tree.
The tree is then passed to HDBSCAN~\cite{hdbscancode} for cluster selection.
Our implementation is capable of handling large sparse hypergraphs, and in our experiments we cluster a million hyperedges in about one minute.

\section{Dataset}\label{sec:data}
\begin{figure}
    \centering
    \includegraphics[width=0.95\linewidth]{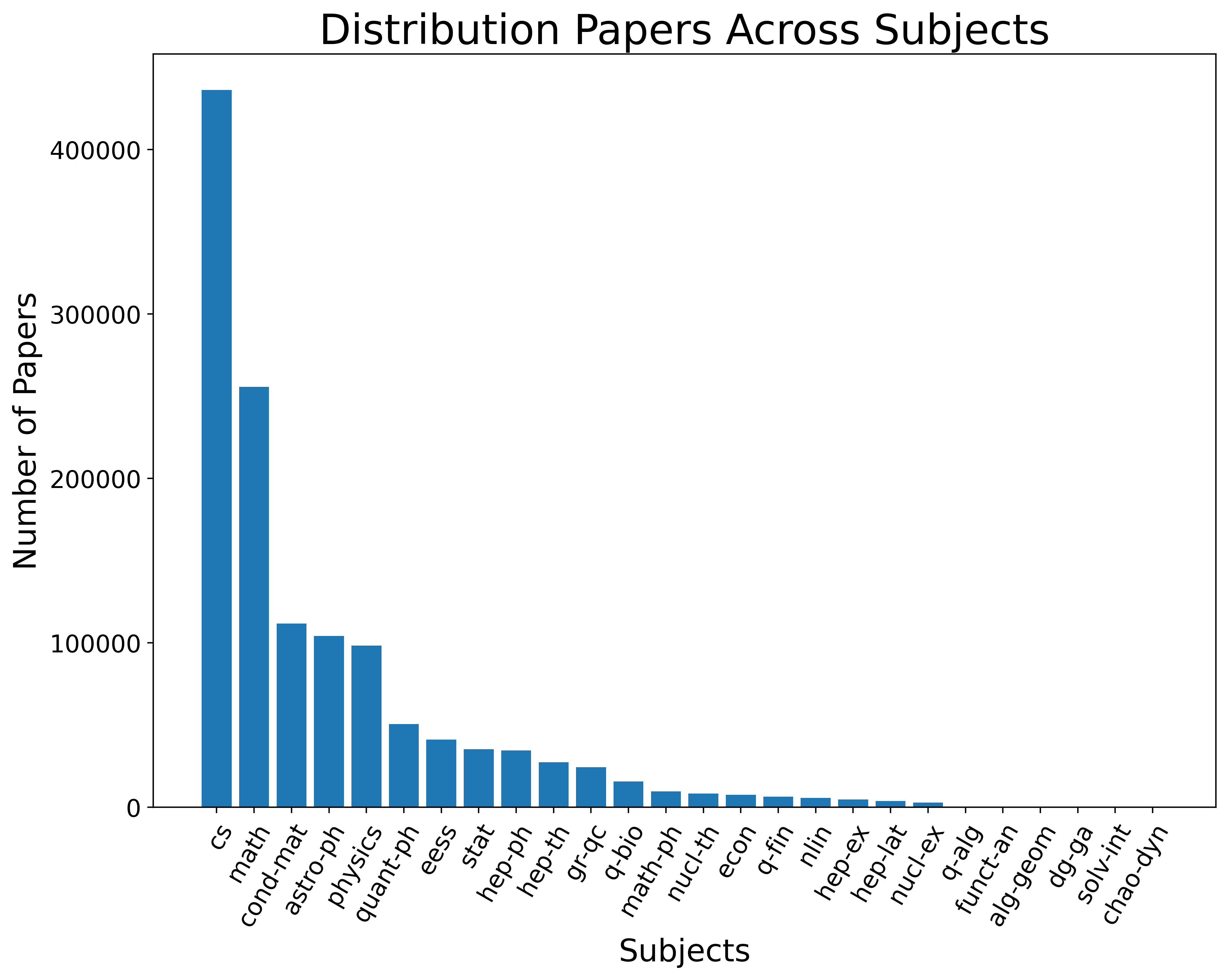}
    \caption{Distribution of primary subjects of the arXiv papers in our dataset.}
    \label{fig:tag_dist}
\end{figure}

For testing our algorithm, we built a collaboration hypernetwork based on papers posted to the arXiv preprint website between January $2018$ and December $2024$~\cite{msadjadi}.
In this temporal hypergraph, vertices correspond to authors and hyperedges correspond to papers.
For a paper $e \in E$, $M(e)$ is the set of authors and $t(e)$ is the time that the paper was uploaded; specifically, $t(e)$ is the number of days after $2018/01/01$.
Each paper is posted with one primary subject and category label: examples are \textit{stat.ml} (Statistics - Machine Learning), \textit{math.at} (Math - Algebraic Topology), and \textit{cs.ai} (Computer Science - Artificial Intelligence).
There are $26$ unique subjects, the distribution of which are shown in Figure \ref{fig:tag_dist}, and $160$ unique subject-category pairs.

During the analysis, we noticed several areas where the construction of the hypergraph introduced errors.
First, each vertex is simply a string of an author's name, so if multiple people have the same name they would be considered as the same vertex.
Furthermore, if a single author publishes under several distinct names (perhaps owing to abbreviations), they will be split into several vertices.
Finally, we noticed that the list of authors is not always actually authors; many of the large physics experiments are listed as authors.
For example: ``n/a the ligo scientific collaboration", ``n/a lhcb collaboration", and ``n/a dune collaboration".
Each of these project-authors seems to start with the string ``n/a", so we removed any authors that match this pattern (and hope there were no real authors inadvertently removed).
This left the graph with $2062$ edges that had no authors, which were also removed.
It is possible that there are other projects listed as authors, but we do not have a systematic method for finding such cases.

\begin{figure}
    \centering
    \includegraphics[height=7.8cm]{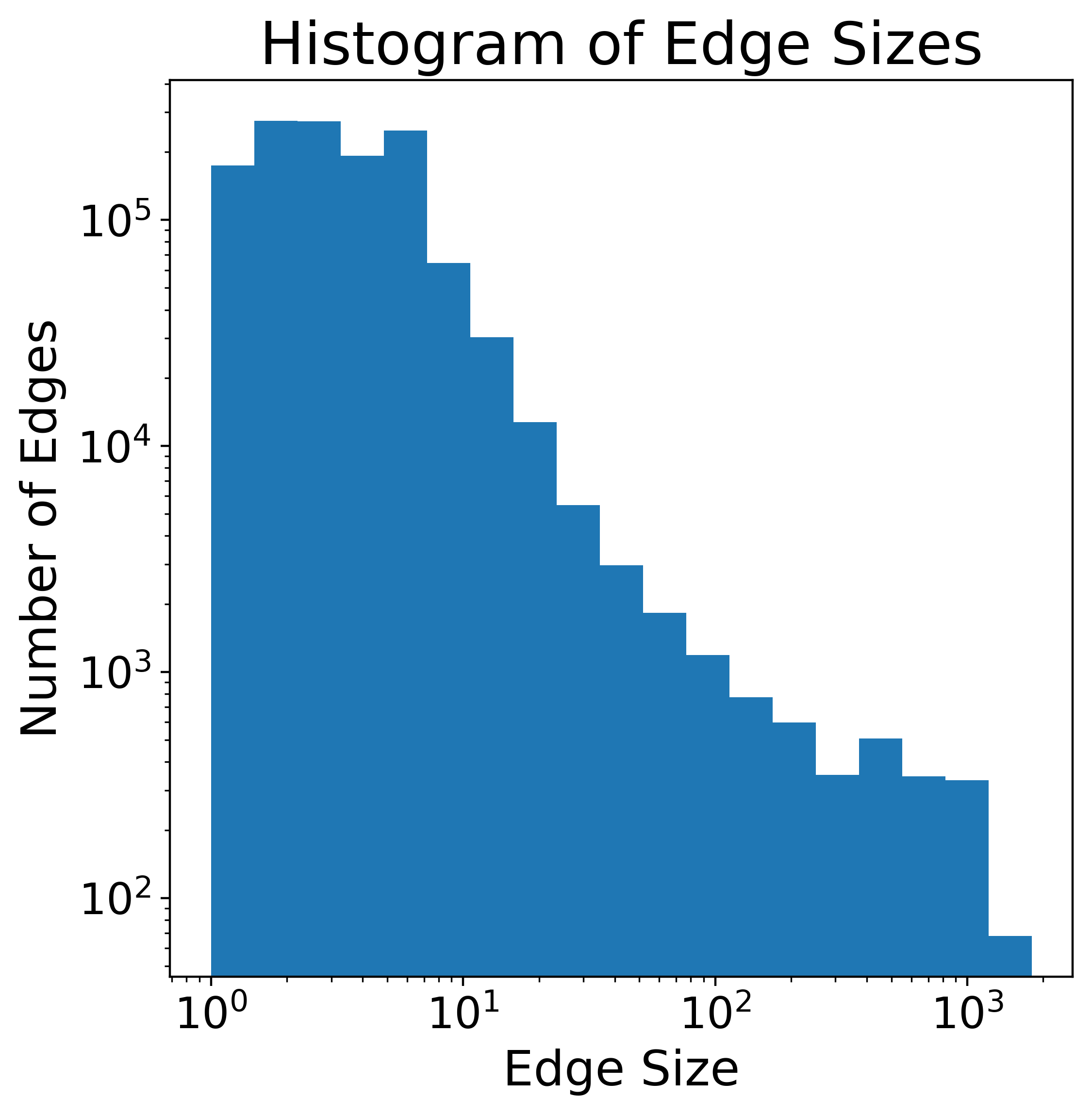}\hfill
    \includegraphics[height=7.8cm]{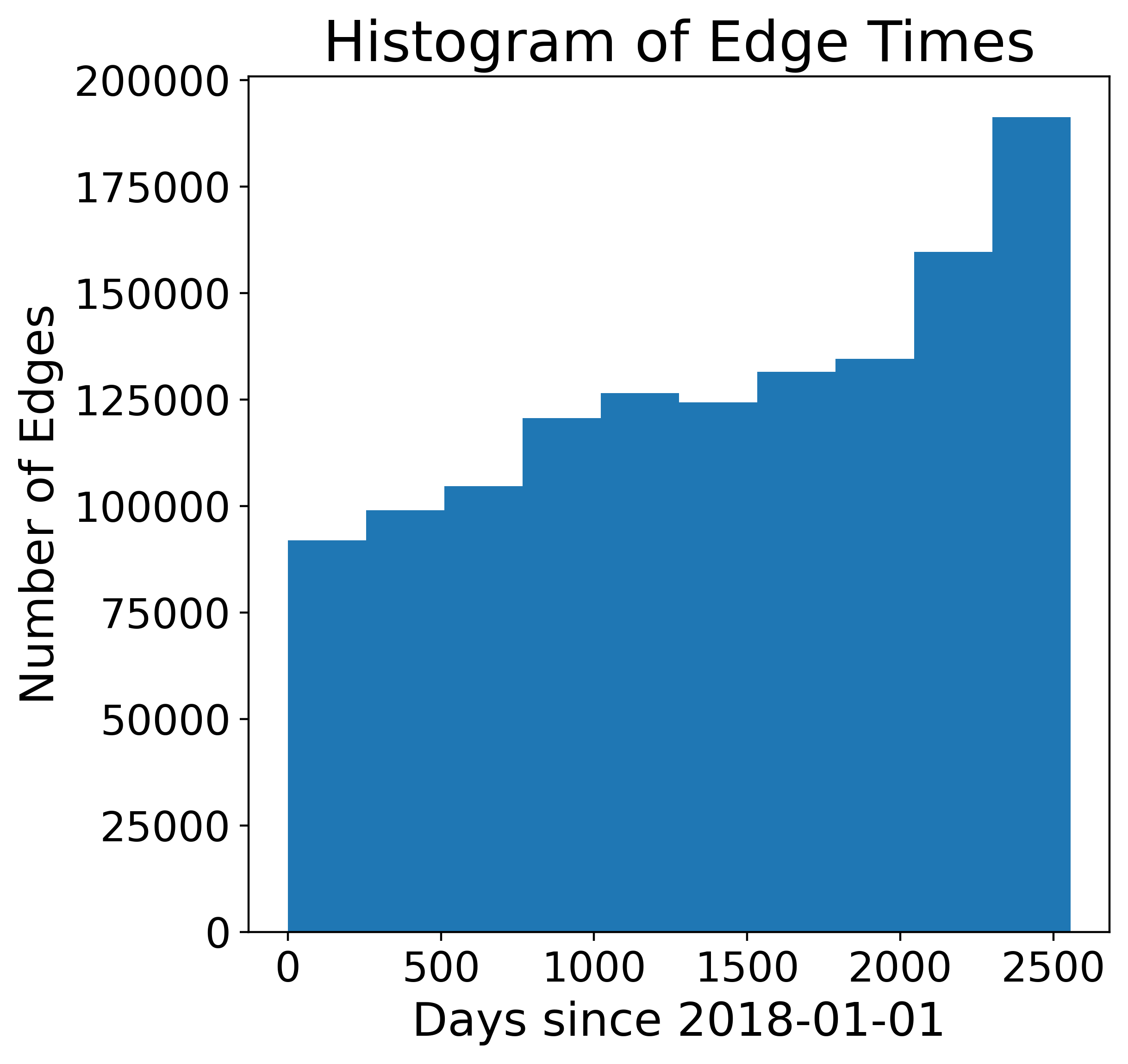}
    \caption{Distribution of the edge sizes (left) and the edge times (right).}
    \label{fig:edge-stats}
\end{figure}

The final temporal hypernetwork has $1,246,592$ nodes and $1,283,931$ hyperedges.
In Figure \ref{fig:edge-stats}, we show the distributions of the edge sizes and the edge timestamps.
The edge sizes appear to follow a power-law distribution.
We also see that the rate at which papers are posted is increasing, but the time period is sufficiently short that the increase does not pose problems when using a single time kernel for every edge. 

\section{Results}\label{sec:results}
\subsection{Effects of the Time Kernel Width} \label{sec:time_kernel_experiments}
\begin{figure}
    \includegraphics[height=7.2cm]{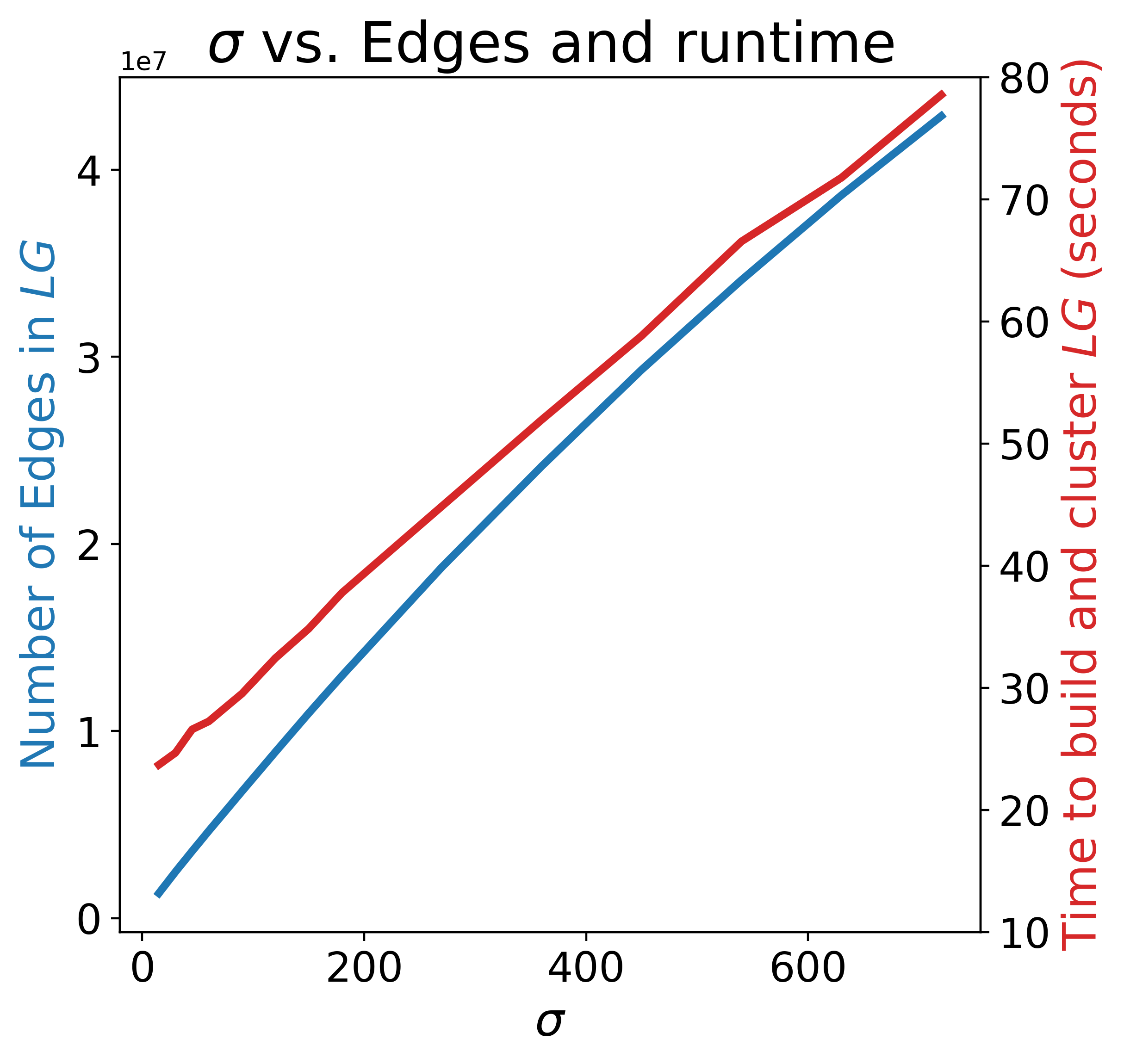}\hfill
    \includegraphics[height=7.2cm]{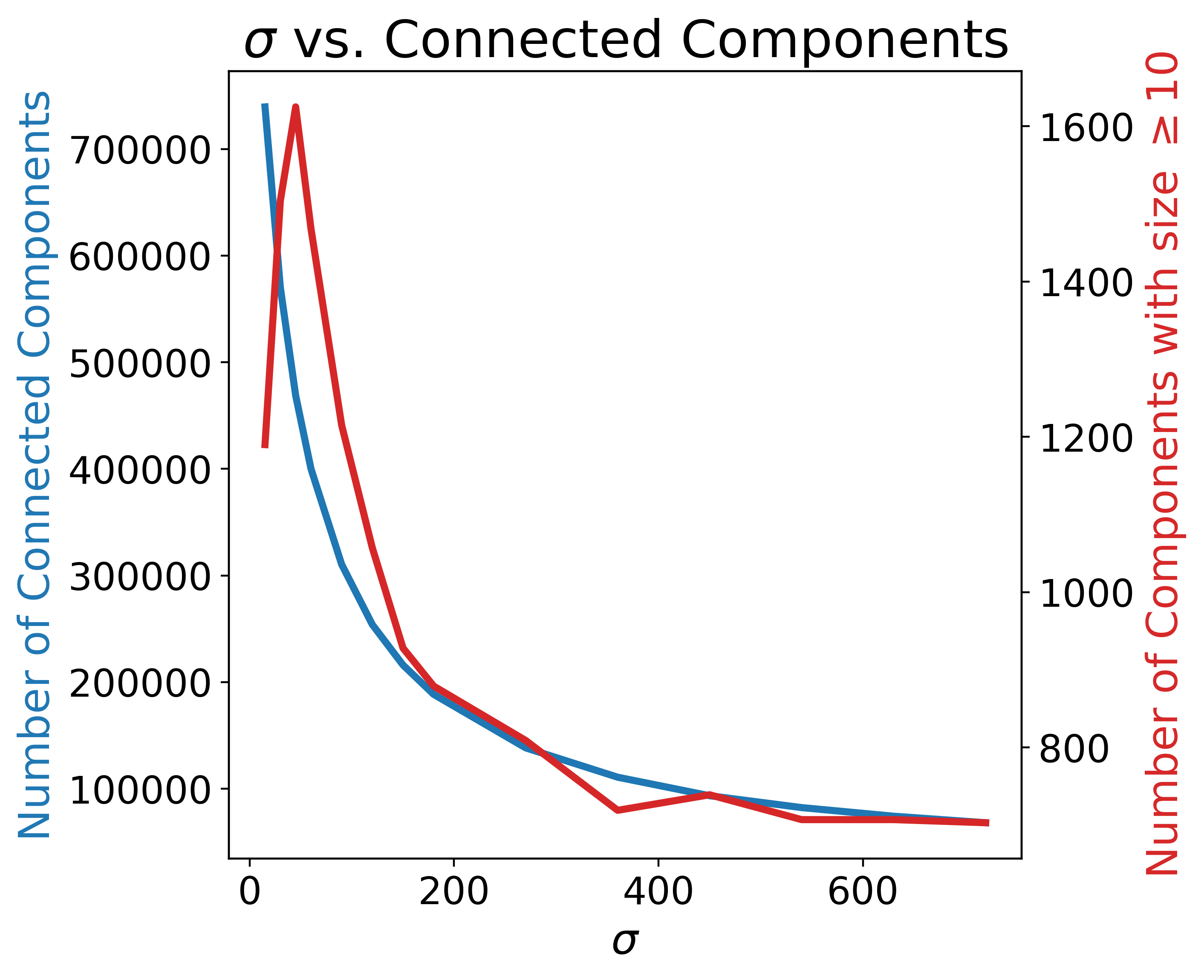}
    \caption{Effect of $\sigma$ on the number of edge in $LG$ and the runtime of the algorithm (left). Effect of $\sigma$ on the number of connected components and large ($\geq 10$) components in $LG$.}
    \label{fig:sigma}
\end{figure}

First, we present some experiments that test the effect of the time kernel parameter $\sigma$.
In Figure \ref{fig:sigma} (left), we show the number of edges in $LG$ and the time to build and cluster $LG$.
Both increase linearly with $\sigma$.
The bottleneck is computing the edge weights for $LG$, so we would expect the time to increase as a linear function of the number of edges in $LG$.
Digging a little deeper, we look at the number of connected components in $LG$ as we vary $\sigma$ (Figure \ref{fig:sigma} right).
We see that the total number of connected components decreases at what looks like an exponential rate.
However, the number of large components (at least $10$ vertices) initially decreases but seems to stabilize when $\sigma$ is around $400$.

\subsection{Comparing a \textit{Narrow} vs. \textit{Wide} Time Kernel}
\begin{figure}
    \centering
    \includegraphics[height=7.8cm]{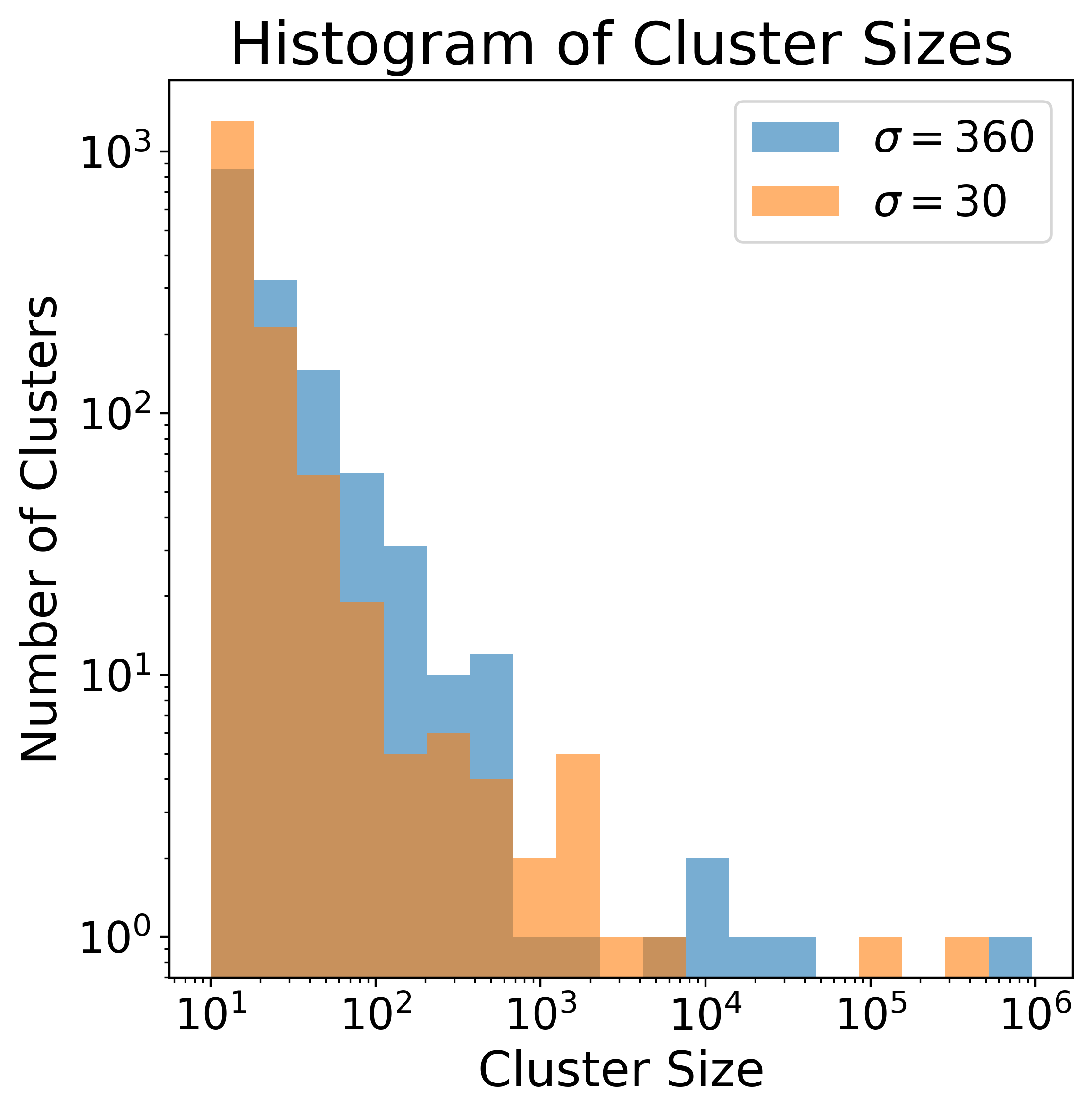}\hfill
    \includegraphics[height=7.8cm]{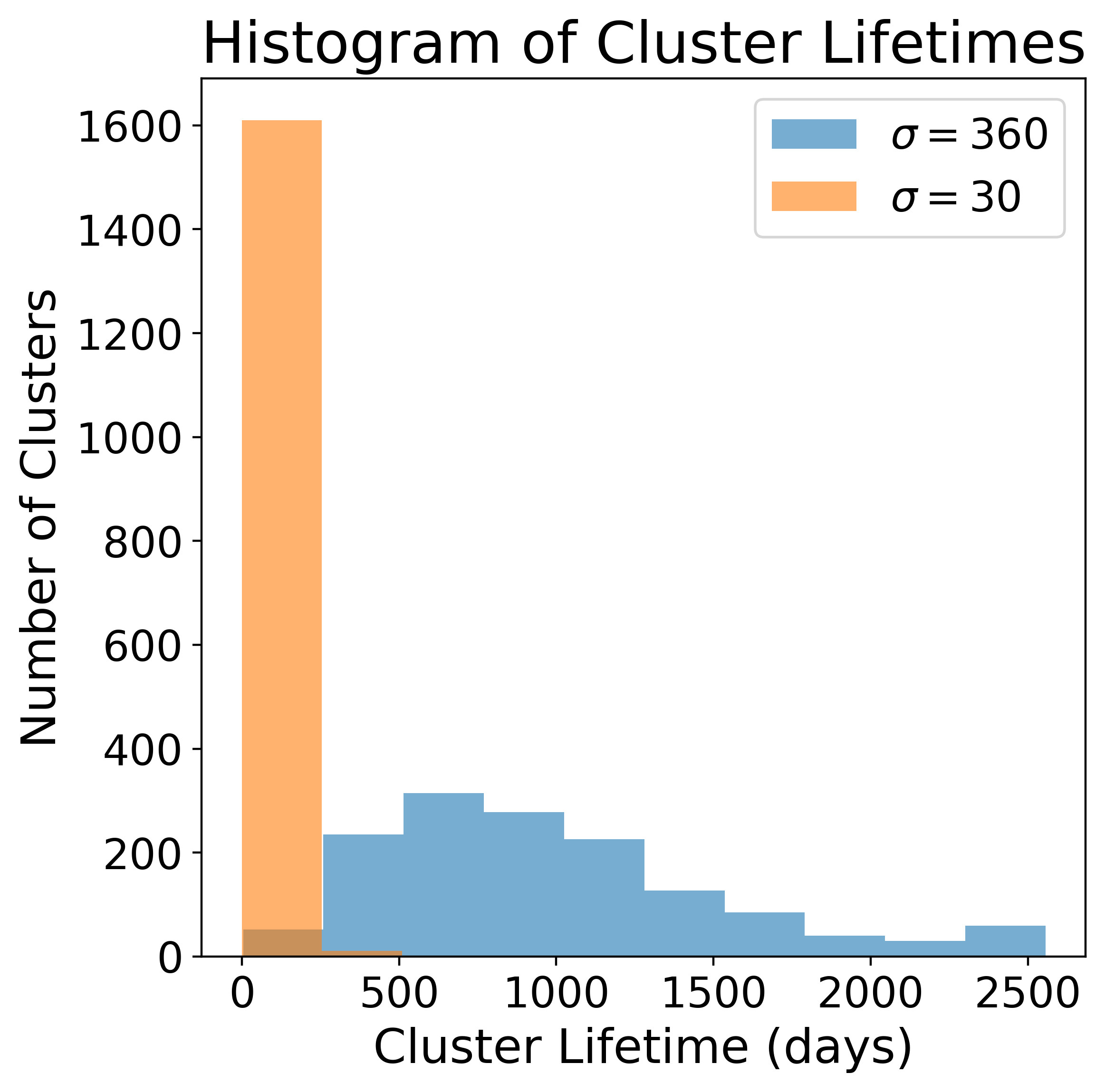}
    \caption{Distribution of the cluster sizes (left) and cluster lifetimes (right) produced by using a narrow ($\sigma = 30$) and wide ($\sigma=360$) time kernel.}
    \label{fig:wide_vs_narrow_dists}
\end{figure}

In this section, we further analyze clusterings produced by a narrow time kernel ($\sigma = 30$) and a wide time kernel ($\sigma = 360$).
The narrow time kernel produced $1,623$ clusters, and $61\%$ of the edges are outliers.
The wide time kernel produced $1,447$ clusters, and  $16\%$ of edges are outliers.
In Figure \ref{fig:wide_vs_narrow_dists}, we compare the distributions of the cluster sizes and cluster lifetimes (the number of days between the first and last paper).
We see that the distribution of cluster sizes are very similar, and both seem to follow an exponential distribution starting at the minimum allowable cluster size of $10$.
However, the distributions of cluster lifetimes look very different.
The wide time kernel produced clusters with a variety of lifetimes, with most clusters having a lifetime of between $1$ and $4$ years, but the maximum is almost the full lifetime of the dataset.
In contrast, almost all the cluster lifetimes from the narrow kernel are very short, usually less than $1$ year.
This behaviour matches our intuition; a group of authors publishing one or more papers every month for a long period of time is very unlikely ($\sigma=30$ means at most $30$ days between papers).

\begin{figure}
    \centering
    \includegraphics[width=0.7\linewidth]{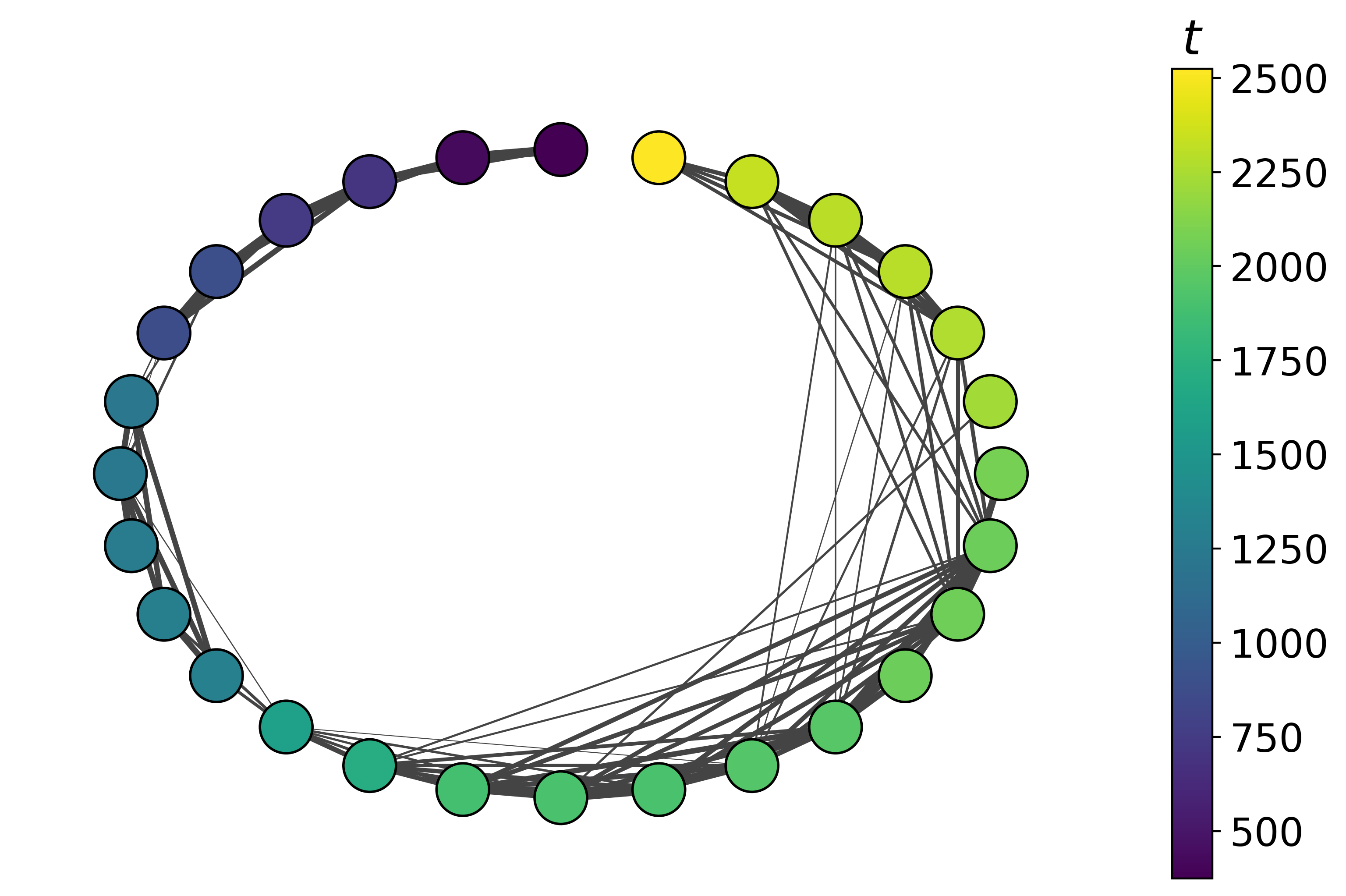}
    \caption{Subgraph of $LG$ for a cluster produced by the wide time kernel. The color of each node corresponds to its timestamp.}
    \label{fig:persistent_cluster}
\end{figure}

Figure \ref{fig:persistent_cluster} shows a subgraph of $LG$ induced by a cluster from the wide kernel clustering that has a long lifetime.
The vertices have been arranged on a circle such that going counter-clockwise around the circle is going forward in time.
This cluster subgraph has a strong ``path-like" structure; each node only links to other nodes that are close in time, so there are no edges through the middle of the circle.
This structure implies a long-term collaboration between authors where they consistently publish together over the course of many years.
The choice of single-linkage clustering in critical to discovering this kind of pattern, since many other clustering techniques would not consider this to be a good cluster because the number of within-cluster edges is not that large. 

\begin{figure}
    \centering
    \includegraphics[height=7.8cm]{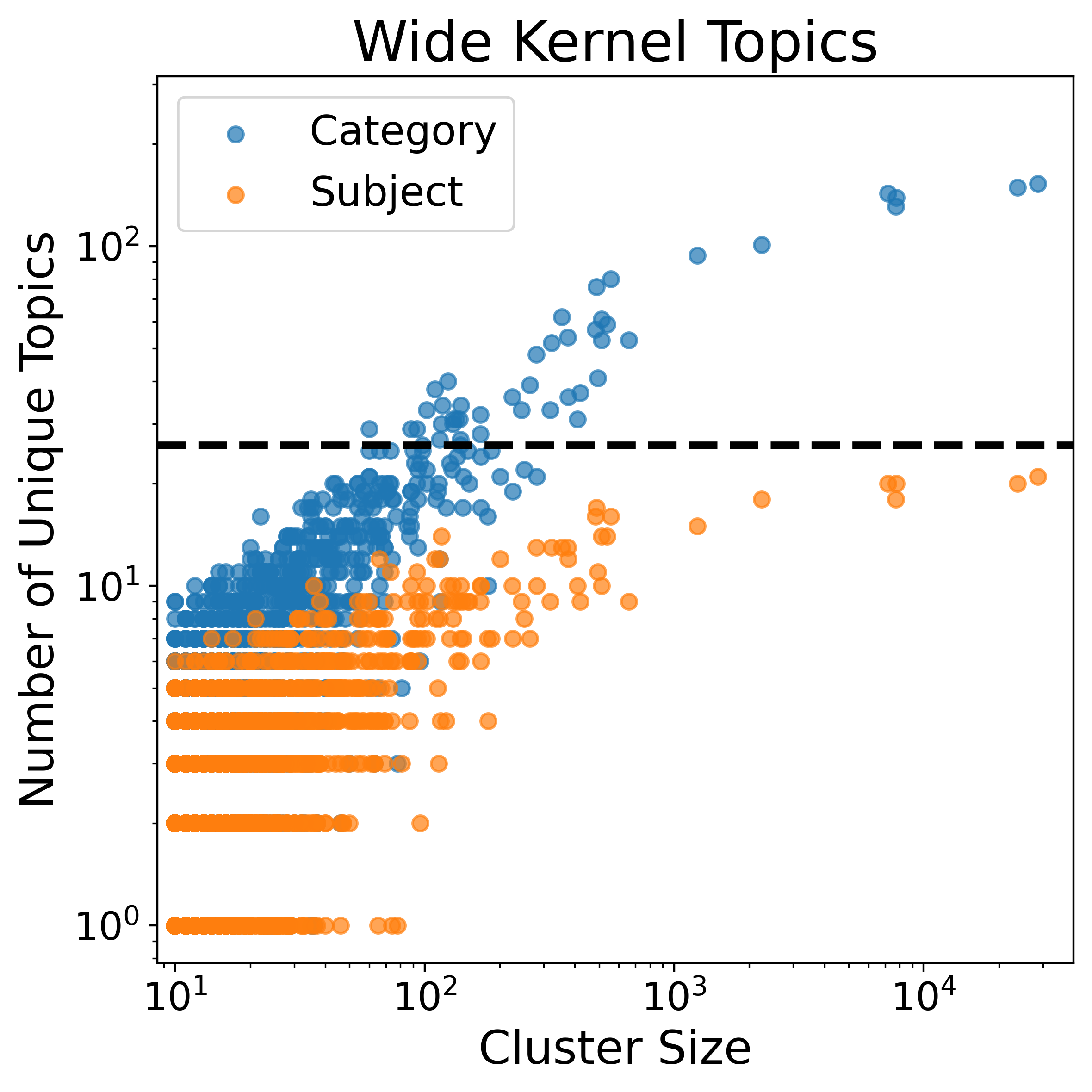} \hfill
    \includegraphics[height=7.8cm]{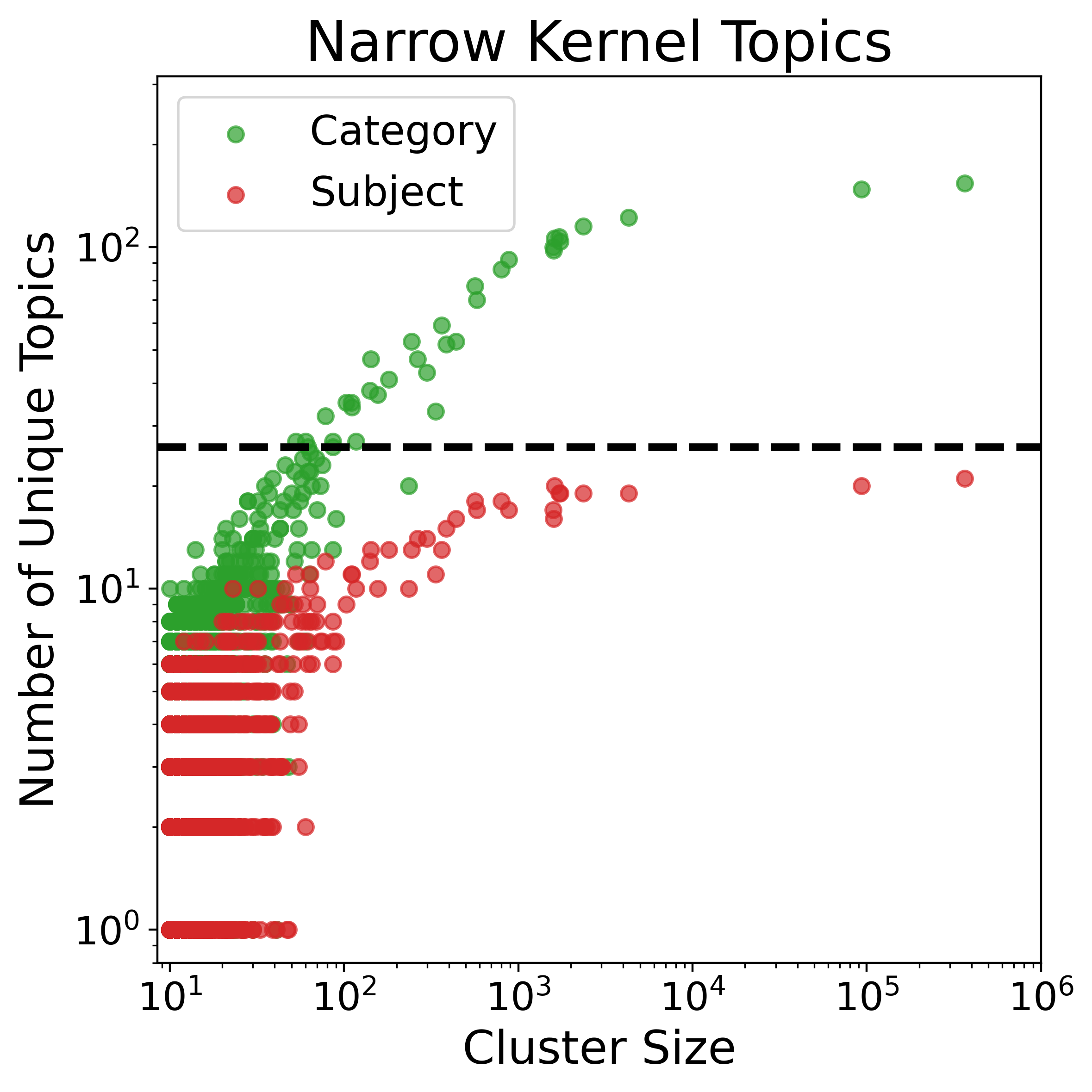}
    \caption{The number of topics in each cluster. We use arXiv tags as a proxy for topic. The horizontal dashed black line represents the number of unique primary tags (subjects) in the dataset.}
    \label{fig:topic_dists}
\end{figure}

In Figure \ref{fig:topic_dists}, we analyze the topic diversity of each cluster.
We consider a subject or category to be a topic, and compare the number of unique topics in each cluster.
Both the wide and narrow kernels produce a positive correlation, with the number of unique subjects often close the maximum number in the dataset for clusters with more than $1000$ papers (the total number in the dataset is represented by the horizontal dashed black line).
In either case, we can conclude that most of the clusters have more than one topic.
We explore concept of cluster topic further in Section \ref{sec:further_analysis}.

\begin{figure}
    \centering
    \includegraphics[height=7.8cm]{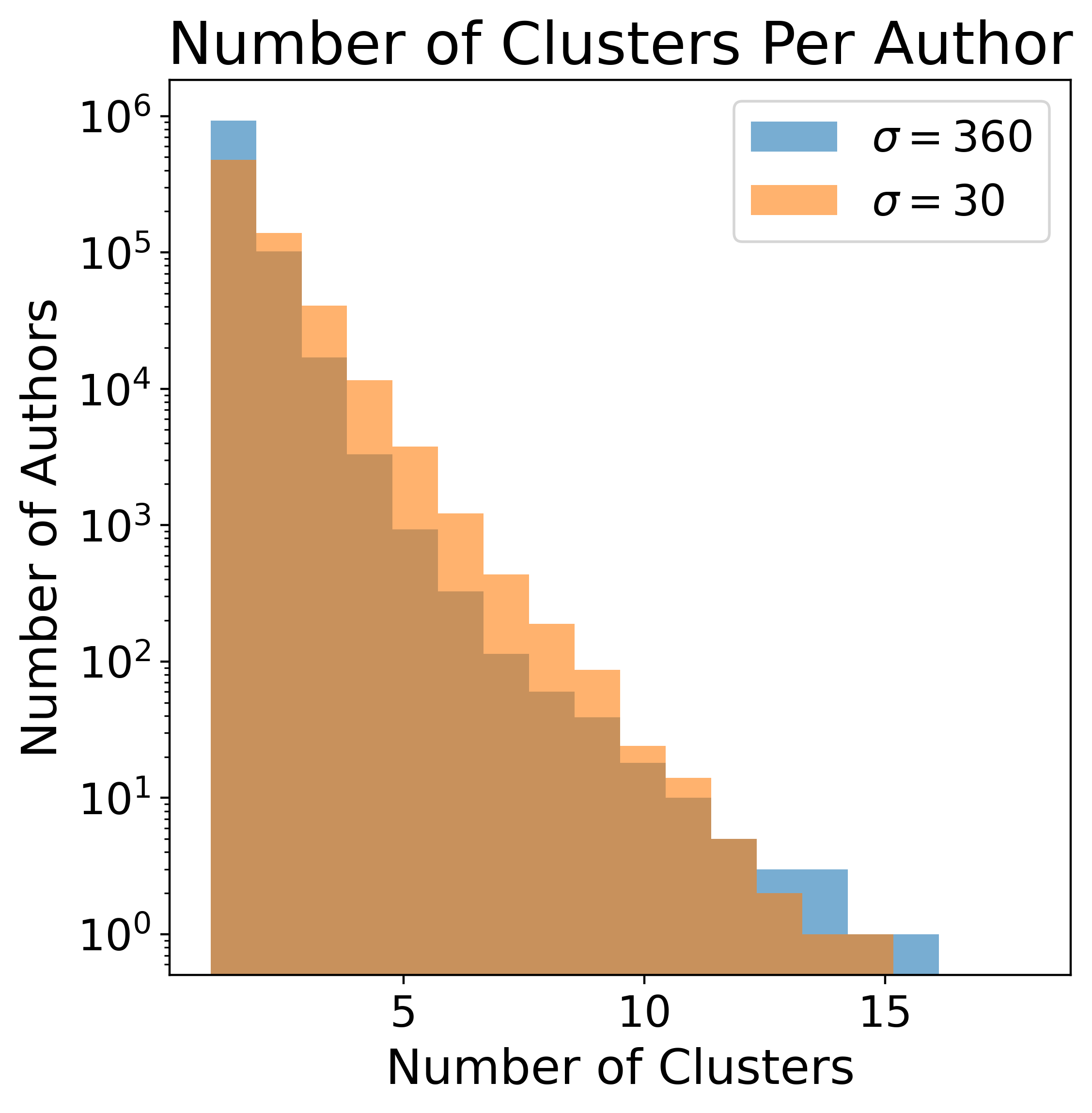} \hfill
    \includegraphics[height=7.8cm]{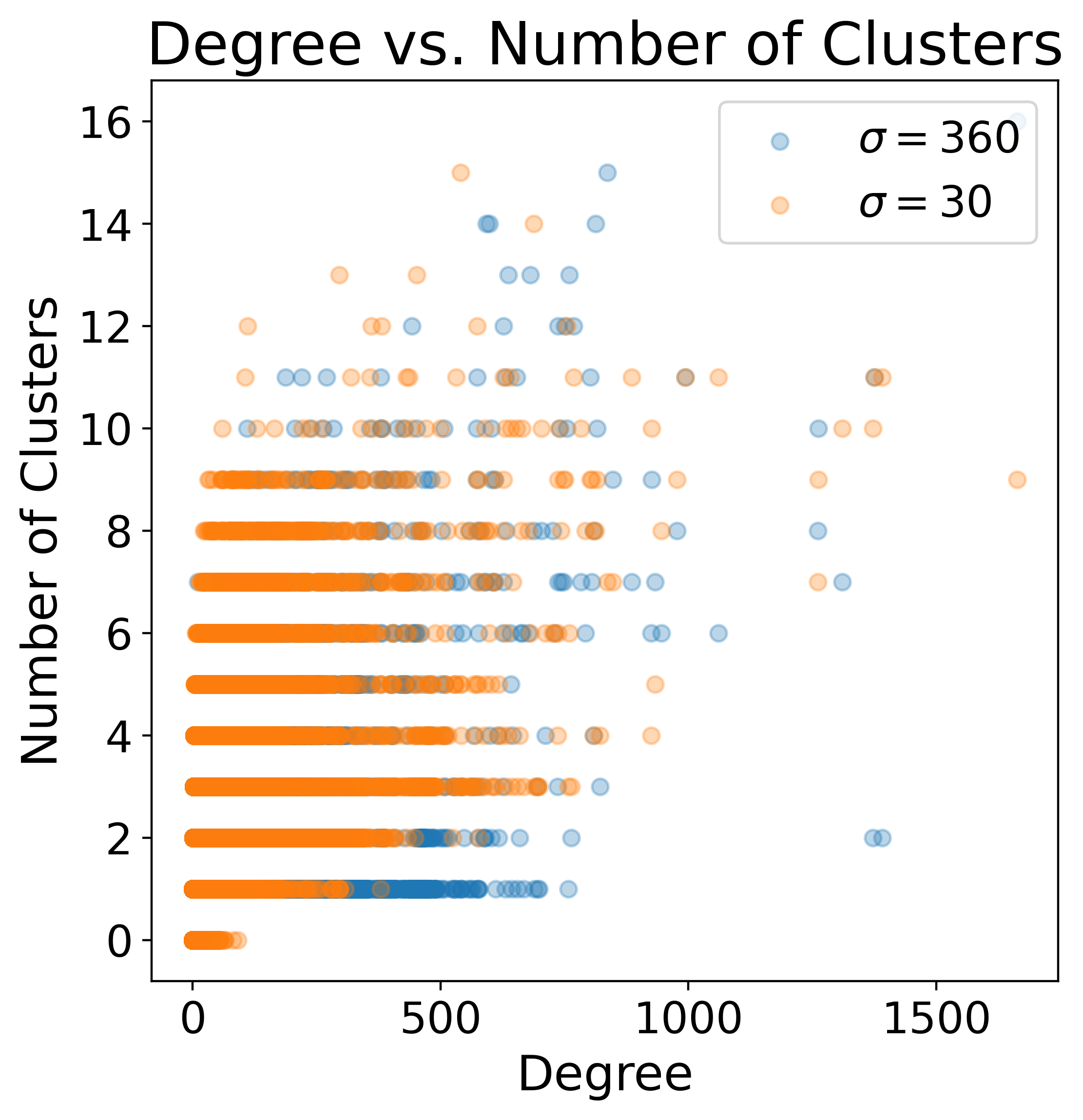}
    \caption{Analysis of projecting the edge clustering to an overlapping clustering of the vertices. We compare the number of cluster per vertex (left) and the number of clusters vs. degree for each vertex (right).}
    \label{fig:author_stats}
\end{figure}

We can also project the edge clustering back to the vertices in the hypergraph, and analyze this overlapping clustering.
If an author has co-authored at least one paper in a cluster, we consider this author as a member of the cluster.
In Figure \ref{fig:author_stats} (left), we show the distribution of the number of clusters per author, and find that both the wide and narrow kernels produce similar power-law looking distributions.
Also, the wide kernel has $16\%$ of authors with zero clusters and the narrow kernel has $46\%$, which is not surprising given the proportions of edges that were classified as outliers.
We also compare the number of papers an author has written with the number of clusters they are part of in Figure \ref{fig:author_stats} (right) and find a weak positive correlation (coefficients of $0.32$ and $0.42$ for wide and narrow respectively). 

\subsection{Further Cluster Analysis} \label{sec:further_analysis}
In this section, we explore two ideas for further analysis of the edge clusters.
For clarity, we will refer to a cluster of edges as a \textit{collaboration}, a fitting name since the set of papers in a cluster would be the result of a sustained collaboration between the authors.

\begin{figure}
    \centering
    \includegraphics[width=0.9\linewidth]{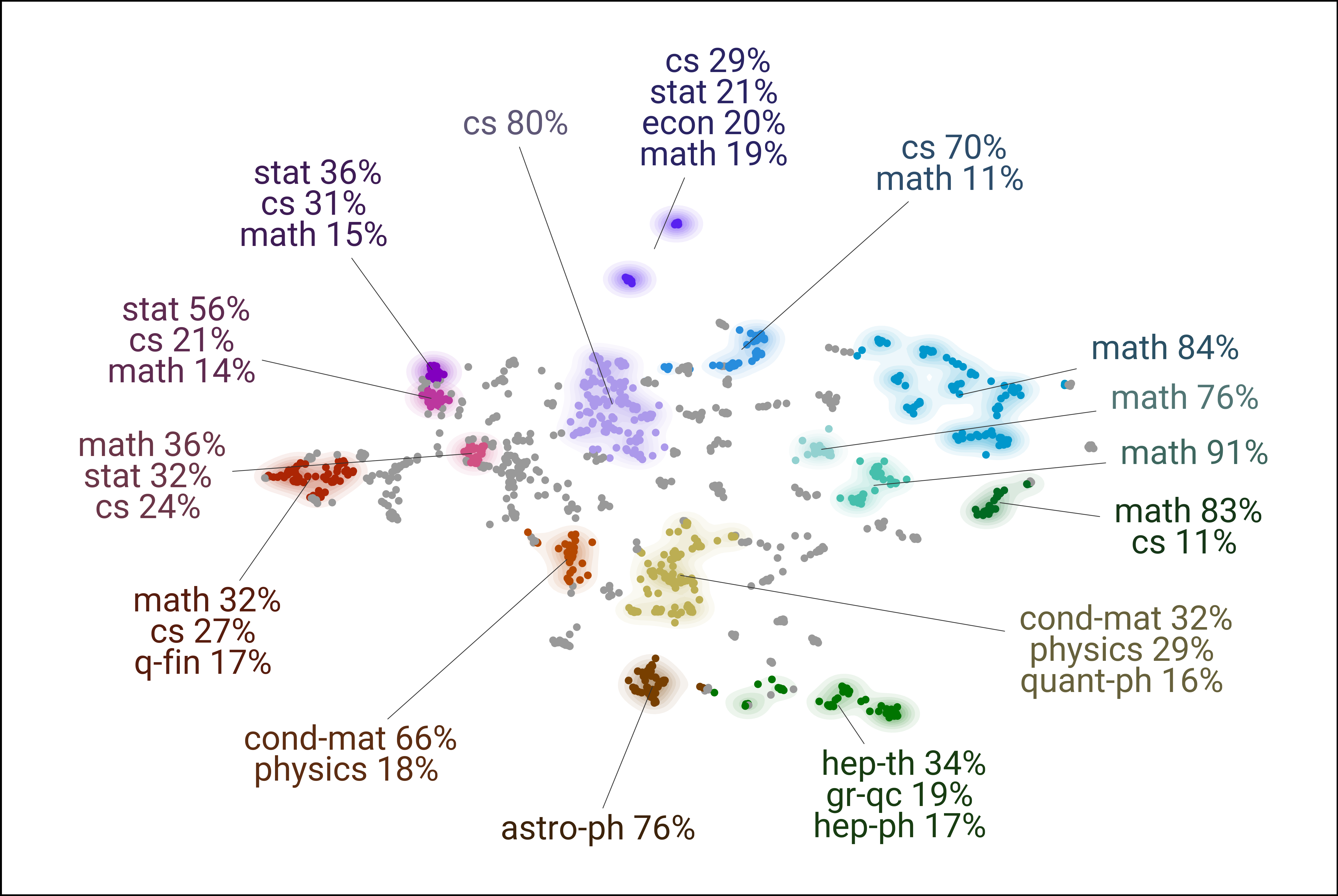}
    \caption{Clusters of collaborations with similar topics distributions. Some clusters are labelled with the proportion of the subjects of papers produced by collaborations in the clusters.}
    \label{fig:topic_clusters}
\end{figure}

First, we look for collaborations from a clustering with $\sigma=360$ that produce papers on a similar set of topics.
For simplicity, we use the primary category as the topic of the paper.
We can consider each collaboration as a \textit{bag of topics}, and using UMAP~\cite{umap} with the Hellinger distance, we can visualize the topic of each collaboration (Figure \ref{fig:topic_clusters}).
We use HDBSCAN~\cite{hdbscancode} to find clusters of collaborations that produced work on similar topics.
Some larger clusters are labelled with the proportion of primary subjects in that cluster (restricted to at least $10\%$ of papers having that primary subject).
There are several clusters of collaborations that are about one subject, like the $76\%$ astro-ph (astrophysics), and the $91\%, 85\%$ and $76\%$ math.
However, many of the clusters have several prominent subjects, like those sharing stat, cs, and math.
Perhaps this is a sign of similarity between the categories in this area so the distinction between the subjects is less clear.
The results of this analysis could be useful for matching groups of authors to predict or encourage future collaborations; if both groups have worked on similar topics before, they are more likely to be interested in similar ideas in the future.

\begin{figure}
    \centering
    \includegraphics[width=0.7\linewidth]{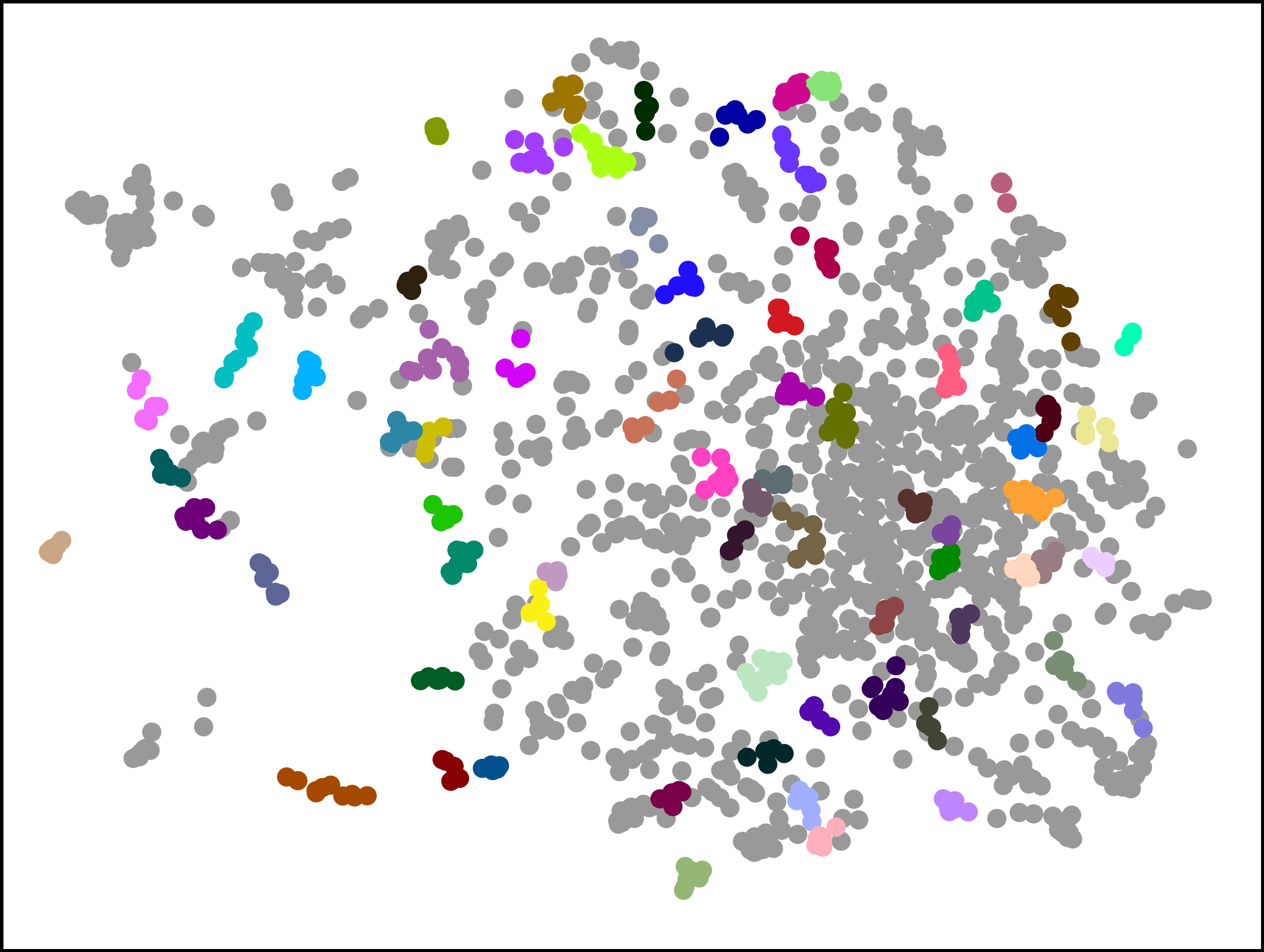}
    \caption{Visualizing clusters from a narrow kernel that have similar author compositions. A few small clusters have been highlighted to represent potential areas of interest for collaborations that disappear and then later reappear.}
    \label{fig:author_comp}
\end{figure}

Second, we can apply a similar pipeline to visualize the similarity in author composition between collaborations.
If we consider each collaboration to be a weighted bag of authors, with weights equal to one divided by the number of authors for that paper, we can again use UMAP and the Hellinger distance to visualize the author composition of each collaboration.
We show the results of this process applied to a  clustering with $\sigma=30$ in Figure \ref{fig:author_comp}.
A potential use case for this visualization is matching collaborations between similar authors that exists during different periods of time.
Dynamic community detection algorithms must contend with communities that disappear and then reappear later, and this analysis could aid in the matching process.
A few clusters found by HDBSCAN having between $5$ and $10$ collaborations that could be investigated in a downstream matching process have been highlighted.

\subsection{Other Structural Similarities}\label{sec:other_s}
In this section, we experiment with two other structural similarity functions to demonstrate the flexibility of the edge clustering framework.

\subsubsection*{Simplicial Pairs}
A recent topic of interest in hypergraphs is simplicial pairs (or encapsulation): a pattern where the vertices of one hyperedge are a subset of another \cite{simplicialpairs, simpliciality, encapsulation}.
For this experiment, we look for \textit{simplicial bursts}, a short time period where there are many simplicial pairs.
We replace the default structural similarity function with an indicator measuring if the two edges are a simplicial pair.
Define this simplicial similarity as
$$s_{Sim}(i, j) = \begin{cases} 1 & M(i) \subseteq M(j) \text{ or } M(j) \subseteq M(i)\\ 0 & \text{otherwise} \end{cases}.$$

\begin{figure}
    \centering
    \includegraphics[height=7.3cm]{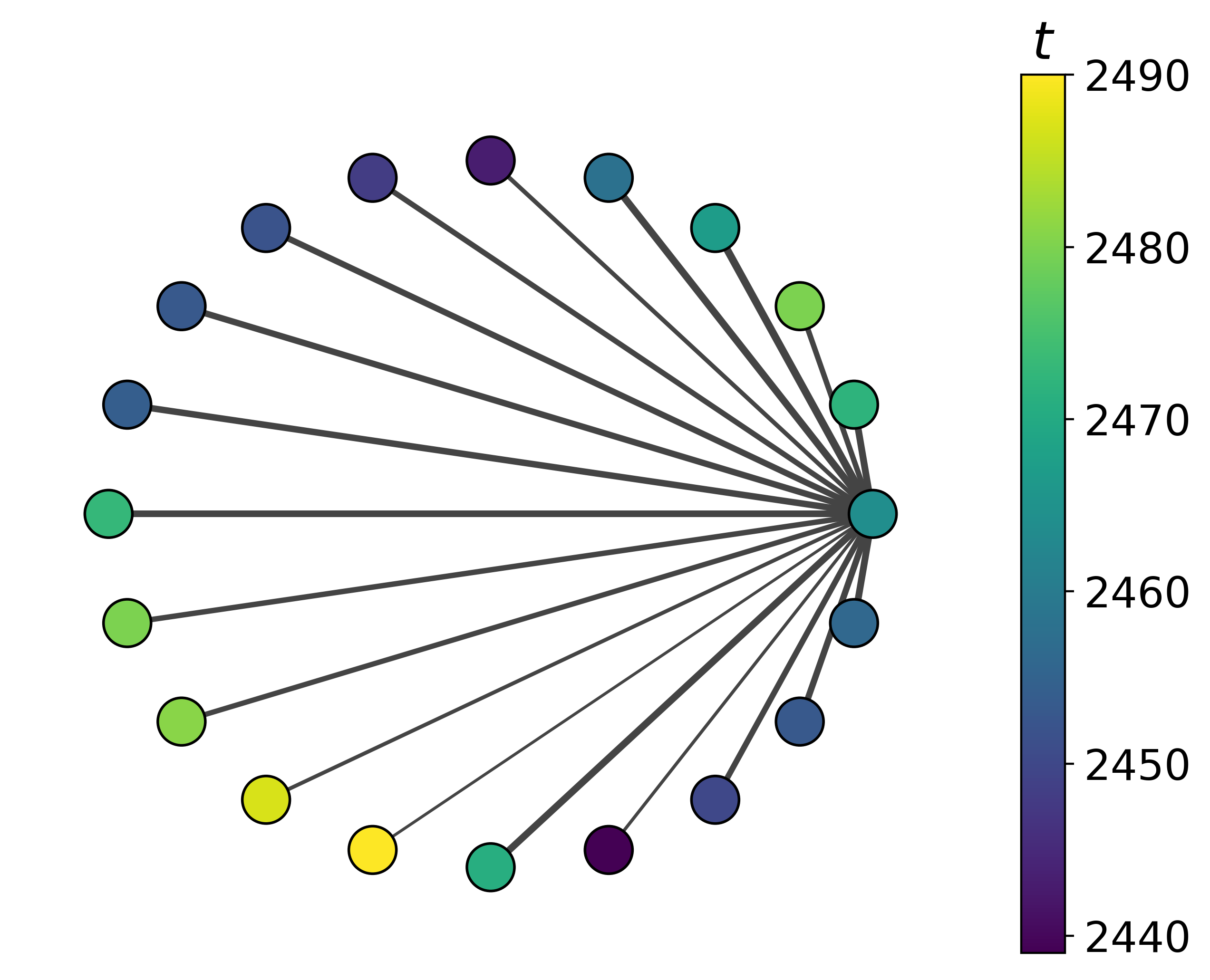} \hfill
    \includegraphics[height=7.3cm]{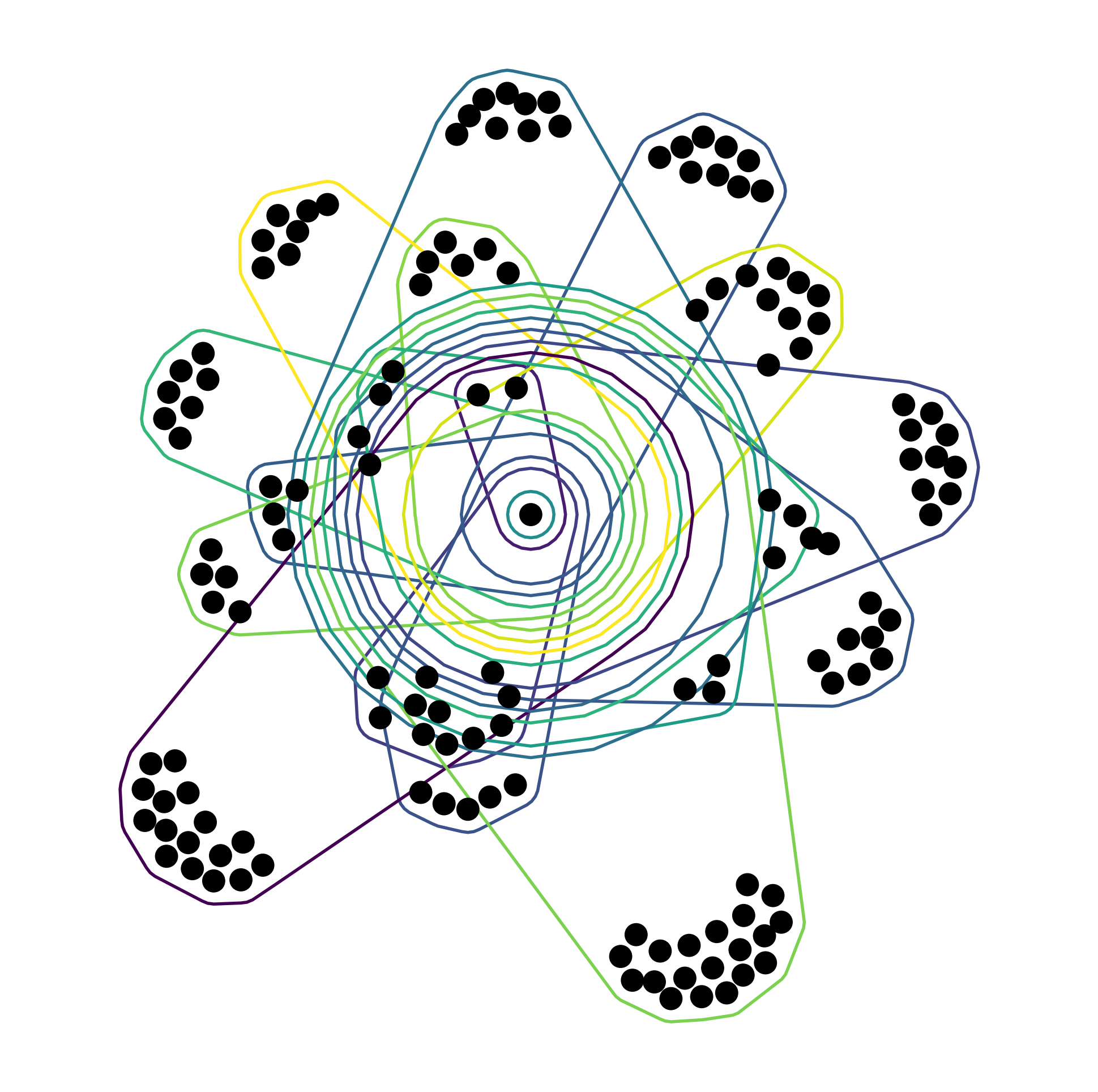}
    \caption{The induced line graph (left) and hypergraph (right) of a typical simplicial burst cluster. The hypergraph visualization was created with Hypernetx~\cite{hypernetx}.}
    \label{fig:simplicial_burst}
\end{figure}

Using $\sigma=30$, we found $609$ clusters with $99\%$ of hyperedges classified as outliers.
As before, the cluster lifetimes are very short, only $73$ days on average and the $95th$ percentile lifetime is only $147$ days.
In Figure \ref{fig:simplicial_burst} we show the line graph and hypergraph induced by a typical cluster.
This cluster has a lifetime of just $51$ days, and we can see that all the simpliciality is driven by the single edge of size one (located in the center of the hypergraph).
The colors of the hyperedges correspond to their timestamp (t), and we can see that the timestamp of the size one edge is approximately halfway through the lifespan of this cluster.

\subsubsection*{Hyperedge Size Filter}
In this experiment, we are focused on finding patterns of interactions that have similar vertices, but also similar sizes.
One could imagine the email hypergraph of a company where vertices are employees and edges are emails that contain the sender and each recipient.
If the company is organized in a hierarchical structure (employees are on a team, teams are in departments, etc.), then we could imagine that there are common email interaction patterns associated to each of these layers (perhaps so common they would be formalized with a mailing list).
To ensure that the team interactions are properly separated from the department interactions, we require that structurally similar edges be of similar size.
We accomplish this by adding an indicator to the Jaccard index as follows: let $\delta(i,j) = \min\{M(i), M(j)\}$ and $\Delta = \max\{M(i), M(j)\}$, then
$$s_{Fil}(i, j) = \begin{cases} s_{Jac}(i,j) & \delta(i,j) * 1.1 + 2 >= \Delta(i,j) \\ 0 & \text{ otherwise } \end{cases}$$
The particular size restriction of at most $110\% + 2$ is somewhat arbitrary, but should suffice to allow for small variations in edge size without allowing for very dissimilar sizes.

\begin{figure}
    \centering
    \includegraphics[width=0.5\linewidth]{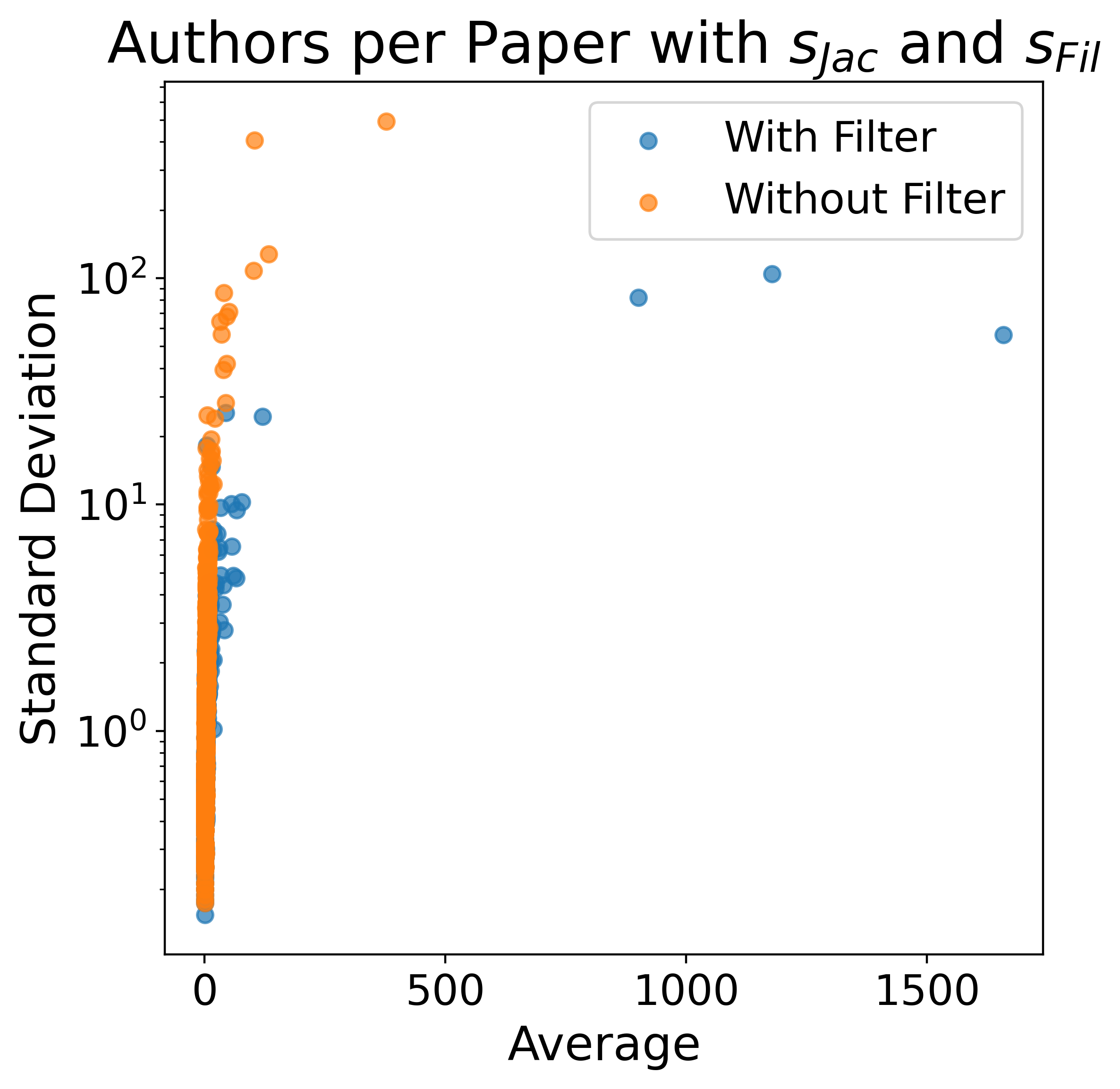}
    \caption{Comparing the average and standard deviation of the number of authors per paper for each cluster.}
    \label{fig:filter_scatter}
\end{figure}

We apply this new structural similarity with $\sigma = 360$.
In Figure \ref{fig:filter_scatter}, we compare the average edge size of each cluster when using $s_{Fil}$ and the previously seen $s_{Jac}$.
The size-filtered similarity function finds several clusters with a much larger average hyperedge size, and the standard deviation of the distribution of edge sizes in these clusters is much lower (note that the y-axis is log scaled).
We investigated the top $5$ clusters with the largest average hyperedge size from the $s_{Fil}$ clustering, and found that each of these clusters corresponds to one or more large physics experiments, which we consider to be a good indication of a real interaction pattern.
Details of the clusters are shown in Table \ref{tab:filtered_clusters}.

\begin{table}
    \centering
    \begin{tabular}{|c|c|c|c|} \hline
         Average Authors Per Paper & \# Papers & Lifetime & Experiment(s) \\ \hline
         $1658.94$ & $35$ & $1275$ & LIGO, Virgo, KAGRA, CHIME/FRB \\
         $1177.82$ & $68$ & $1877$ & LIGO, Virgo, DES, KAGRA, DUNE \\
         $900.84$ & $231$ & $1875$ & LHCb, DUNE, ANTARES \\
         $120.30$ & $67$ & $1289$ & Fermi-LAT, LISA \\
         $77.45$ & $20$ & $711$ & KOPPARAPU, SoLid \\
         \hline
    \end{tabular}
    \caption{Details of the $5$ clusters with the largest average number of authors per paper found using the $s_{Fil}$ structural similarity function. Each of the clusters corresponds to two or more large physics experiments.}
    \label{tab:filtered_clusters}
\end{table}

\section{Future Work}\label{sec:conclusion}
As seen in the variety of experiments in this paper, approaching edge clustering with the line graph is a very flexible approach for community detection.
In this section, we suggest directions for improvement and further experimentation.

We proposed a few structural similarity measures, but the most useful choice would likely be problem specific, and we would be interested in which ones are useful for different types of networks.
If we allow $LG$ to be a hypergraph, we could even consider patterns of interaction that require more than two edges, like those in the anti-money laundering literature \cite{anitmoneylaundering}, although this would require clustering a hypergraph.
The edge similarity measure could also consider other information, such as the abstract when edges correspond to papers.

In addition, we demonstrated that the choice of time kernel scale $\sigma$ has a considerable impact on the clustering.
We are interested in applying a normalization to the time kernel $T_\sigma$, as was done in \cite{dbmapper} for vector data, to account for different time densities in different parts of $LG$.

Combining potentially both of the above ideas, we are interested in applications of \textit{temporally aware random edge walks}.
The edge clustering algorithm in ~\cite{edgeclustering1} derived their edge similarity measure from random walks, and Aksoy et al. \cite{hypernetworkwalks} extended several measures based on random walks to hypergraphs.
HEdge2Vec~\cite{hedge2vec} is a recent  application random edge walks on a hypergraph; it is a method to embed the hyperedges as vectors and is similar to the very popular node2vec~\cite{node2vec} algorithm.
Creating temporally aware hyperedge embeddings would be helpful for intuitively visualizing an edge clustering.

Finally, our algorithm relies on a graph community detection algorithm, and while single linkage has been effective in many algorithms, further explorations of graph clustering algorithms that allow for outliers and overlap \cite{cas}, or detection of ``path-like" clusters would be beneficial.

\bibliography{references}

\end{document}